\documentclass[showpacs,twocolumn]{revtex4}
\usepackage{graphicx,amsmath}

\begin{document}

\title{Dynamical polarization, screening, and plasmons in gapped graphene}

\author{P.K. Pyatkovskiy}

\affiliation{Kyiv National Taras Shevchenko University, Kyiv, 03127,
Ukraine}

\begin{abstract}
The one-loop polarization function of graphene has been calculated at zero temperature for arbitrary wavevector, frequency, chemical potential (doping), and band gap. The result is expressed in terms of elementary functions and is used to find the dispersion of the plasmon mode and the static screening within the random phase approximation. At long wavelengths the usual square root behaviour of plasmon spectra for two-dimensional (2D) systems is obtained. The presence of a small (compared to a chemical potential) gap leads to the appearance of a new undamped plasmon mode. At greater values of the gap this mode merges with the long-wavelength one, and vanishes when the Fermi level enters the gap. The screening of charged impurities at large distances differs from that in gapless graphene by slower decay of Friedel oscillations ($1/r^2$ instead of $1/r^3$), similarly to conventional 2D systems.
\end{abstract}


\maketitle

\section{Introduction}
Graphene, a free-standing flat single layer of carbon atoms, has attracted considerable attention since its experimental discovery~\cite{Novoselov2004S} due to its unique electronic properties. The conduction and the valence bands in graphene touch each other at zero energy at two inequivalent points, $K$ and $K'$, in the momentum space. Close to those points charge carriers have a linear isotropic energy spectrum $E_\pm({\bf k})=\pm\hbar v_F|{\bf k}|$~\cite{Wallace1947PR} and behave like relativistic massless Dirac particles in ${2+1}$ dimensions, with the Fermi velocity $v_F\simeq10^6$~m/s playing the role of the speed of light~\cite{DiVincenzo1984PRB}. This remarkable feature makes graphene different from all other known two-dimensional systems with their usual quadratic dispersion law of charged quasiparticles. For the undoped graphene at zero temperature the valence band is completely filled (`Dirac sea') and the conduction band is empty, which corresponds to a zero value of the chemical potential (Fermi energy) $\mu$. The latter can be made non-zero and regulated experimentally by the applied external electric field~\cite{Novoselov2004S}

One of the most important fundamental quantities for understanding physical properties of graphene is the dynamical polarization function $\Pi(\omega,k)$ which describes the screening of the Coulomb potential due to many-body effects and determines the collective excitation modes. This function for graphene at finite chemical potential and $\omega=0$ has been calculated in the one-loop approximation in~\cite{Gorbar2002PRB}. Later, in~\cite{Wunsch2006NJP} and~\cite{Hwang2007PRB} it was obtained for the arbitrary frequency, and the results were used, in particular, to find the dispersion of the plasmon mode within the random phase approximation (RPA).

The present paper deals with the more general case, when a finite gap $2\Delta$ between the conduction and the valence band exists in graphene, and the charge carriers behave analogously to relativistic massive Dirac particles, with the dispersion relation $E_\pm({\bf k})=\pm\sqrt{\hbar^2v_F^2k^2+\Delta^2}$. Such a gap is observed in experiments with graphene placed on the silicon carbide~\cite{Zhou2007NM} or graphite~\cite{Li2008} substrate (with the band gap $2\Delta$ of about 0.26~eV or 10~meV, respectively), and is commonly attributed to the sublattice symmetry breaking due to the commensurate perturbation from the substrate. Moreover, even in the absence of any substrate, there should still remain a small gap ($2\Delta_{so}\sim10^{-3}$~meV) originating from the spin-orbit interaction~\cite{Huertas-Hernando2006PRB}. The gap can also be generated dynamically by applying an external magnetic field~\cite{Gorbar2002PRB,Khveshchenko2001PRL}. The dispersion of a plasmon mode for graphene with the finite spin-orbit fermionic gap has been obtained numerically in~\cite{Wang2007PRB}. Our aim is to explore in more detail the influence of the gap on the plasmon dispersion and the static charge screening. In this work we restrict ourselves to the case of zero temperature and calculate analytically the dynamical polarization function for gapped graphene. Next we find the screened Coulomb potential at large distances and the spectrum of plasmons for different values of the dimensionless parameter $\Delta/\mu$ by solving the dispersion equation with the obtained polarization function.

\section{Model}
The low-energy Hamiltonian of graphene can be written in a Dirac-like form~\cite{Kane2005PRL},
\begin{equation}
\begin{split}
{\cal H}_0&=-i\hbar v_F\sum_{s=\pm}\Psi^\dag_s(\tau_3\otimes\sigma_1\partial_1+\tau_0\otimes\sigma_2\partial_2)\Psi_s\\
&=-i\hbar v_F\sum_{s=\pm}\bar\Psi_s(\gamma^1\partial_1+\gamma^2\partial_2)\Psi_s\,,
\label{hamiltonian}
\end{split}
\end{equation}
where $\Psi_s=(\psi^s_{KA},\psi^s_{KB},\psi^s_{K'A},\psi^s_{K'B})^T$ is a four-component wavefunction of charged quasiparticles, the components of which refer to the $A$ and $B$ sublattices and to two inequivalent valleys in momentum space (around $K$ and $K'$ points). Pauli matrices $\sigma_i$ and $\tau_i$ act on sublattice and valley components, respectively ($\sigma_0$, $\tau_0$ are unit matrices), $s$ is a spin index, gamma matrices are defined as $\gamma^0=\tau_0\otimes\sigma_3$, $\gamma^1=i\tau_3\otimes\sigma_2$, $\gamma^2=-i\tau_0\otimes\sigma_1$, and $\bar\Psi_s\equiv\Psi^\dag_s\gamma^0$ is the Dirac conjugated spinor.

Interaction with a substrate of a certain type may break the sublattice symmetry. In this case the Hamiltonian~(\ref{hamiltonian}) acquires an additional term
\begin{equation}
{\cal H}_\Delta=\Delta\sum_{s=\pm}\Psi^\dagger_s\tau_0\otimes\sigma_3\Psi_s=\Delta\sum_{s=\pm}\bar\Psi_s\Psi_s\,,
\label{diracgap}
\end{equation}
which is equivalent to the usual Dirac mass term (in the current representation of $\gamma$-matrices), and the energy gap $2\Delta$ opens between the valence and the conduction band. Another gap is driven by a spin-orbit interaction which adds the term
\begin{equation}
{\cal H}_{so}=\Delta_{so}\sum_{s=\pm}s\Psi^\dagger_s\tau_3\otimes\sigma_3\Psi_s
\label{spinorbitgap}
\end{equation}
to the Hamiltonian.

\section{Polarization function}
The screening of an instantaneous bare Coulomb interaction
\begin{equation}
U_0(t,r)=\frac{e^2\delta(t)}{r}=e^2\int\frac{d\omega}{2\pi}\int\frac{d^2k}{2\pi}\frac{\exp(i{\bf kr}-i\omega t)}{k}
\end{equation}
due to many-body effects is determined by the retarded polarization function $\Pi(\omega,k)$ or, equivalently, by the dielectric function $\varepsilon(\omega,k)\equiv1+\Pi(\omega,k)/k$, resulting in
\begin{equation}
U(t,r)=\frac{e^2}{\varepsilon_0}\int\frac{d\omega}{2\pi}\int\frac{d^2k}{2\pi}\frac{\exp(i{\bf kr}-i\omega t)}{k+\Pi(\omega,k)}\,,
\end{equation}
where $\varepsilon_0$ is an effective background dielectric constant, the average of that of the substrate and that of the vacuum. Note that $\Pi(\omega,k)$ is defined here in the same way as in~\cite{Gorbar2002PRB}, which differs from the polarization function used in~\cite{Wunsch2006NJP,Hwang2007PRB,Wang2007PRB} by a factor of $-2\pi e^2/\varepsilon_0$.

We consider clean graphene, i.e. the effect of scattering of charge carriers on impurities is not taken into account. The polarization function is proportional (with a factor of $2\pi/\varepsilon_0$) to the time component of the vacuum polarization tensor in a three-dimensional quantum electrodynamics (QED$_3$), where $v_F$ is the `speed of light' and $\Delta/v_{F}^2$ is a fermion mass. We calculate $\Pi(\omega,k)$ analytically in the one-loop approximation. The exact results are given in the appendix by the equation~(\ref{pizeromu}) for the case $|\mu|<\Delta$ and by the equations~(\ref{repi})--(\ref{impi}) for $\mu>\Delta$. In what follows it is assumed that $\mu>0$ and $\omega>0$, taking into account that the polarization function depends only on the absolute value of $\mu$ and $\Pi(-\omega,k)=\Pi(\omega,k)^*$.

At small energies and momenta $\hbar v_Fk\ll\hbar\omega\ll\mu$ and $\mu>\Delta$ it is easy to obtain from~(\ref{repi}) and~(\ref{impi})
\begin{equation}
\Pi(\omega,k)\simeq-\frac{e^2N_fk^2\mu}{\varepsilon_0\hbar^2\omega^2}\biggl(1-\frac{\Delta^2}{\mu^2}\biggr)\,.
\label{pismallk}
\end{equation}
Here $N_f$ is the number of fermions in QED$_3$ (in the case of graphene $N_f=2$ due to the spin degeneracy).

In the absence of a band gap the expression for the polarization function~(\ref{repi})--(\ref{impi}) reduces to the one calculated in~\cite{Wunsch2006NJP} and~\cite{Hwang2007PRB}, and can be written in the following way
\begin{equation}
\begin{split}
&\Pi(\omega,k)=\frac{2e^2N_f\mu}{\varepsilon_0\hbar^2v_F^2}-\frac{e^2N_fk^2}{4\varepsilon_0\hbar\sqrt{v_F^2k^2-\omega^2}}\\
&\times\biggl[\frac{2\mu+\hbar\omega}{\hbar v_Fk}\sqrt{1-\Bigl(\frac{2\mu+\hbar\omega}{\hbar v_Fk}\Bigr)^2}+i{\,\rm arccosh}\Bigl(\frac{2\mu+\hbar\omega}{\hbar v_Fk}\Bigr)\\
&+\frac{2\mu-\hbar\omega}{\hbar v_Fk}\sqrt{1-\Bigl(\frac{2\mu-\hbar\omega}{\hbar v_Fk}\Bigr)^2}-i{\,\rm arccosh}\Bigl(\frac{2\mu-\hbar\omega}{\hbar v_Fk}\Bigr)\biggr]\,,
\label{pizerodelta}
\end{split}
\end{equation}
with the prescription $\omega\to\omega+i0$. This expression is an analytic function of $\omega$ without singularities in the whole upper complex half-plane. Hence it coincides also with the result obtained in~\cite{Barlas2007PRL} when calculated on the positive imaginary half-axis.

\section{Static screening}
In the RPA approximation, the static screening of the Coulomb potential $\phi_0(r)=Ze/r$ of an impurity with charge density $n_0(r)=Ze\delta({\bf r})$ is determined by the static ($\omega=0$) one-loop polarization function. The induced charge density in the graphene plane reads
\begin{equation}
\delta n(r)=Ze\int\frac{d^2k}{(2\pi)^2}\biggl[\frac{1}{\varepsilon_0\varepsilon(0,k)}-1\biggr]\exp(i{\bf kr})\,,
\label{inddensity}
\end{equation}
and the resulting screened potential is
\begin{equation}
\phi(r)=\frac{Ze}{\varepsilon_0}\int\frac{d^2k}{2\pi}
\frac{\exp(i{\bf kr})}{k+\Pi(0,k)}\,. \label{scrpotential}
\end{equation}
At zero frequency and $\mu>\Delta$ we recover from~(\ref{repi})--(\ref{impi}) the previous result for the static polarization function~\cite{Gorbar2002PRB}
\begin{equation}
\begin{split}
\Pi(0,&k)=\frac{2e^2N_f\mu}{\varepsilon_0\hbar^2v_F^2}\biggl[1-\theta(k-2k_F)\biggl(\frac{\sqrt{k^2-4k_F^2}}{2k}\\
&-\frac{\hbar^2v_F^2k^2-4\Delta^2}{4\hbar v_Fk\mu}{\,\rm arctan\,}\frac{\hbar v_F\sqrt{k^2-4k_F^2}}{2\mu}\biggr)\biggr]\,.
\label{pizerofrequency}
\end{split}
\end{equation}
At large distances from the charged impurity ($k_Fr\gg1$), there are two main contributions to~(\ref{inddensity}) and~(\ref{scrpotential}). The first, Thomas-Fermi contribution, which is determined by the long-wavelength ($k\to0$) behavior of the polarization function,
\begin{equation}
\Pi(0,k<2k_F)=\frac{2e^2N_f\mu}{\varepsilon_0\hbar^2v_F^2}\,,
\end{equation}
is the same as in gapless graphene~\cite{Wunsch2006NJP}
\begin{equation}
\delta
n_{TF}(r)=-\frac{e^2N_f\mu}{\pi\varepsilon_0\hbar^2v_F^2}\phi_{TF}(r)=-\frac{Z\hbar^2v_F^2}{4\pi N_fe\mu r^3}\,,
\end{equation}
and decays under the same $1/r^3$ law as in an ordinary non-relativistic two-dimensional electron gas (2DEG)~\cite{Stern1967PRL}. The second contribution is oscillatory (Friedel oscillations) and comes from the non-analyticity of the polarization function~(\ref{pizerofrequency}) at $k=2k_F$ where its derivative is discontinuous (for $\Delta\ne0$). These oscillations
\begin{equation}
\delta n_{osc}(r)=\frac{k_F}{\pi}\phi_{osc}(r)=-\frac{Ze^3N_f\hbar^2v_F^2k_F\Delta^2\sin(2k_Fr)}{2\pi\mu\bigl(\varepsilon_0\hbar^2v_F^2k_F+e^2N_f\mu\bigr)^2r^2}\,,
\end{equation}
decay as $1/r^2$ as in a 2DEG~\cite{Stern1967PRL} and slower than in gapless graphene where they obey the $1/r^3$ law~\cite{Cheianov2006PRL,Wunsch2006NJP}. The reason for this difference is that at $\Delta=0$ the polarization function at $k=2k_F$ has a discontinuity only in the second derivative, in contrast to the case of the gapped graphene and 2DEG.

At $\mu<\Delta$, the static limit of~(\ref{pizeromu}) reads
\begin{equation}
\begin{split}
\Pi(0,k)={}&\frac{e^2N_f}{2\varepsilon_0\hbar^2v_F^2}\Biggl(2\Delta+\frac{\hbar^2v_F^2k^2-4\Delta^2}{\hbar v_Fk}\\
&\times\arcsin\frac{\hbar v_Fk}{\sqrt{\hbar^2v_F^2k^2+4\Delta^2}}\Biggr)
\label{pizerofreqmu}
\end{split}
\end{equation}
(compare with equation (14) in~\cite{Gorbar2002PRB}), and the induced charge density~(\ref{inddensity}) at large distances, which is determined by the long-wavelength behaviour of~(\ref{pizerofreqmu}),
\begin{equation}
\Pi(0,k\to0)\simeq\frac{e^2N_fk^2}{3\varepsilon_0\Delta}\,,
\end{equation}
decays as $1/r^3$~\cite{Kotov2008PRB},
\begin{equation}
\delta n(r)\simeq\frac{Ze^3N_f}{6\pi\varepsilon_0^2\Delta r^3}\,.
\end{equation}
The potential at large distances for this case
\begin{equation}
\phi(r)\simeq\frac{Ze}{\varepsilon_0r}\biggl(1-\frac{e^4N_f^2}{9\varepsilon_0^2\Delta^2r^2}\biggr)\,,
\end{equation}
in the main order remains screened only by the substrate.

\section{Plasmon dispersion}
The plasmon dispersion $\omega_p(k)$ is obtained within the RPA approximation by finding zeros of the dielectric function $\varepsilon(\omega,k)$ or, equivalently, by solving the equation
\begin{equation}
\Pi(\omega_p,k)+k=0\,.
\label{dispeqn}
\end{equation}
Using the approximate expression~(\ref{pismallk}) we obtain immediately for small $\omega$ and $k$
\begin{equation}
\omega_p(k)\simeq\sqrt{\frac{e^2N_fk\mu}{\varepsilon_0\hbar^2}\biggl(1-\frac{\Delta^2}{\mu^2}\biggr)}\,,
\label{dispsmallk}
\end{equation}
with a square root behaviour usual for two-dimensional systems~\cite{Stern1967PRL}, which differs from the case of gapless graphene~\cite{Wunsch2006NJP,Hwang2007PRB} only by the factor $\sqrt{1-\Delta^2/\mu^2}$.

To obtain the plasmon dispersion for an arbitrary frequency and wavevector, the equation~(\ref{dispeqn}) with the polarization function defined at~(\ref{repi}) and~(\ref{impi}) is solved numerically. We use the following parameters: $\varepsilon_0=2.5$ (this value corresponds to the case when a graphene sheet is deposed on a $\rm SiO_2$ substrate) and $v_F=10^6$~m/s.

\begin{figure*}
\includegraphics[width=6cm]{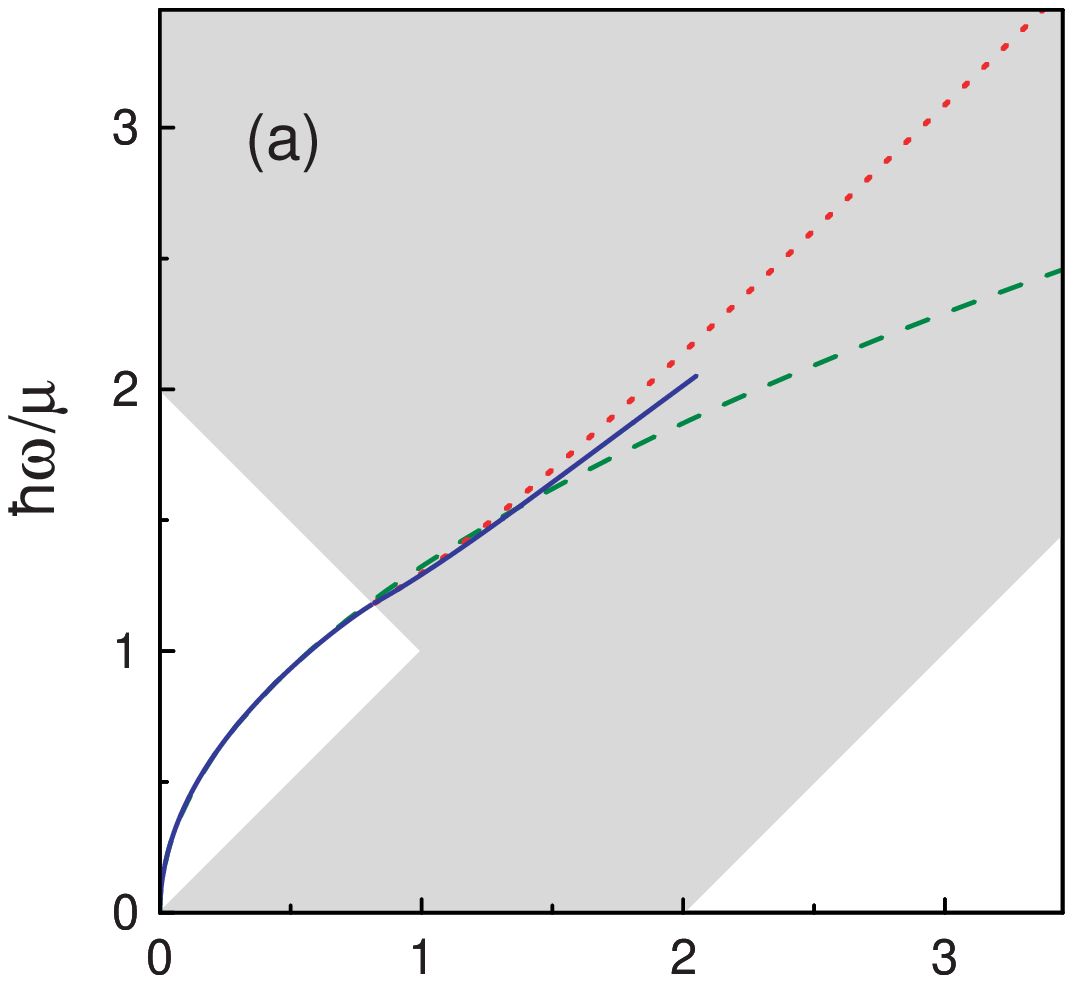}\hspace{0.3cm}\includegraphics[width=6cm]{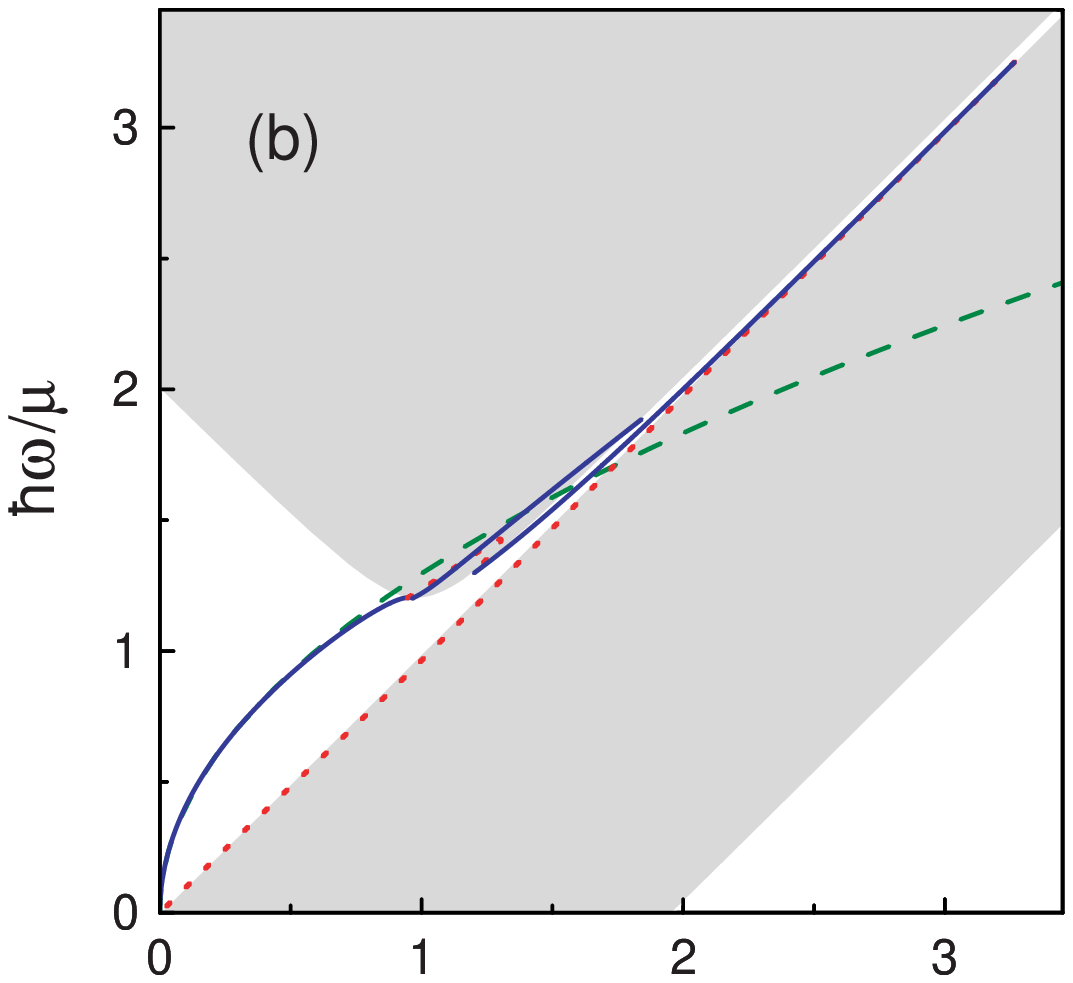}
\includegraphics[width=6cm]{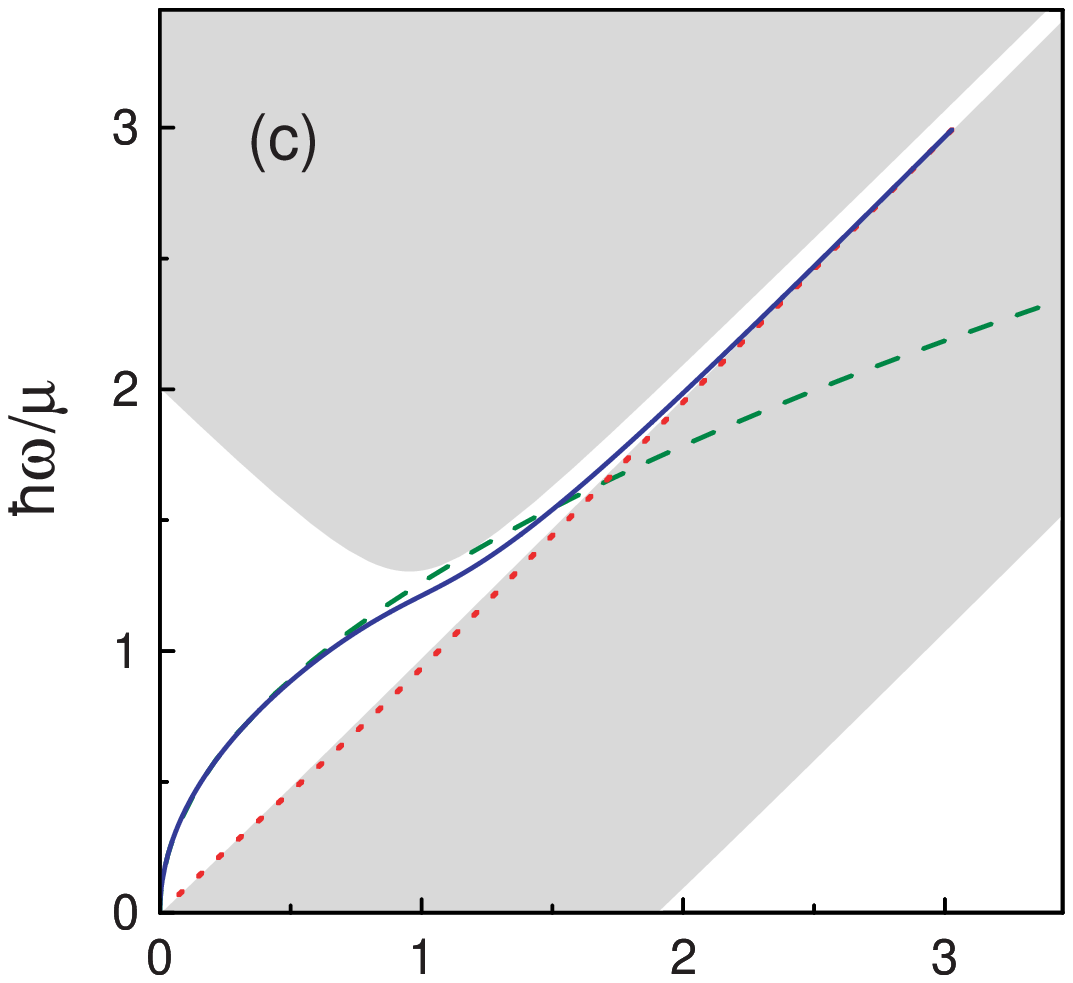}\hspace{0.3cm}\includegraphics[width=6cm]{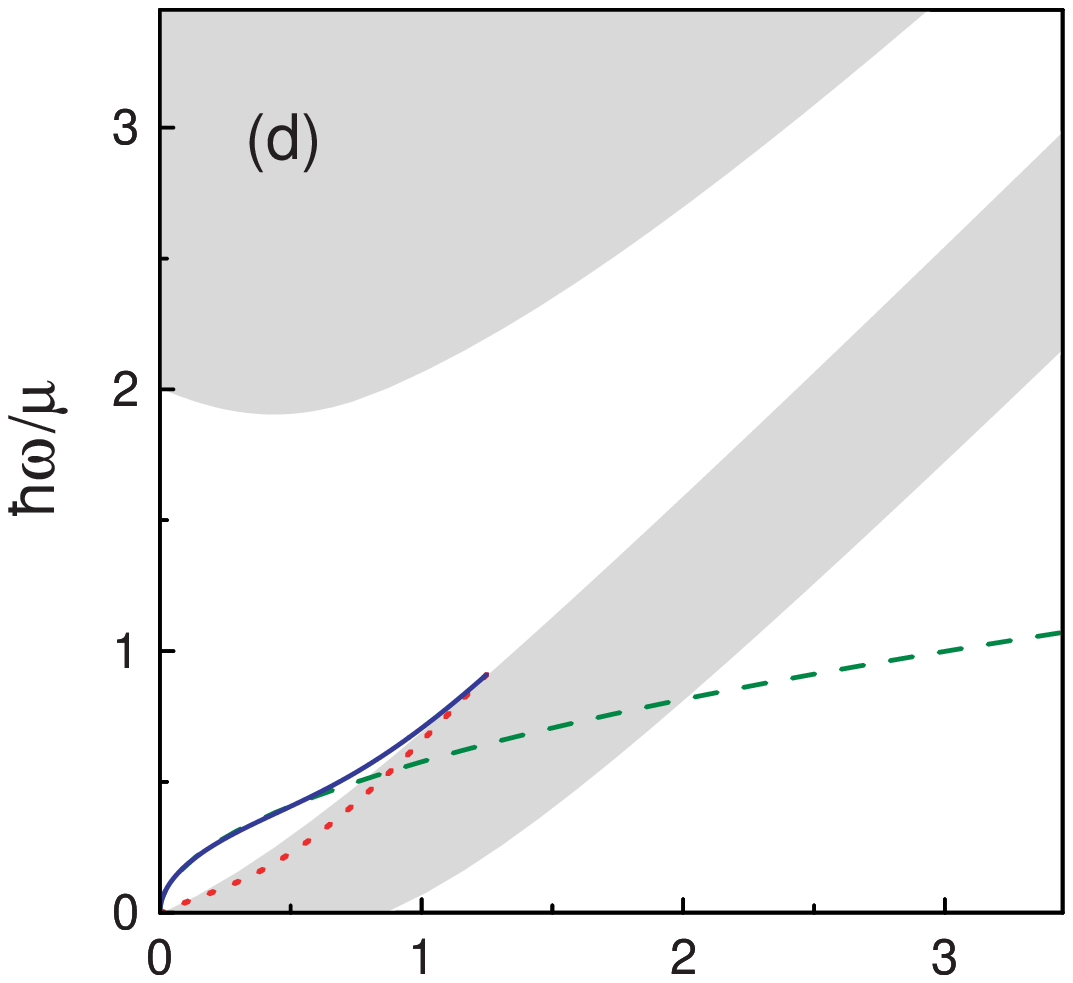}
\includegraphics[width=6cm]{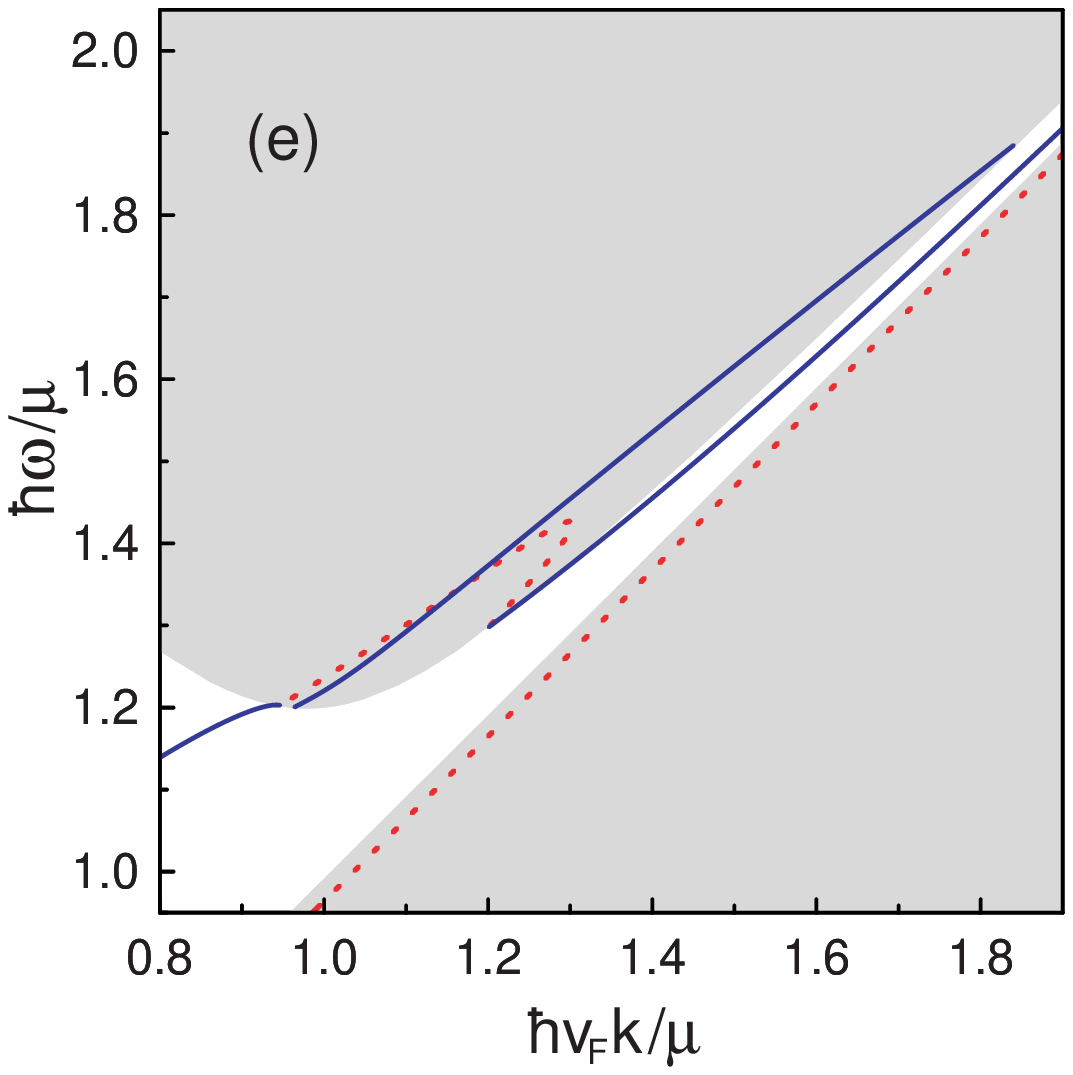}\hspace{0.3cm}\includegraphics[width=6cm]{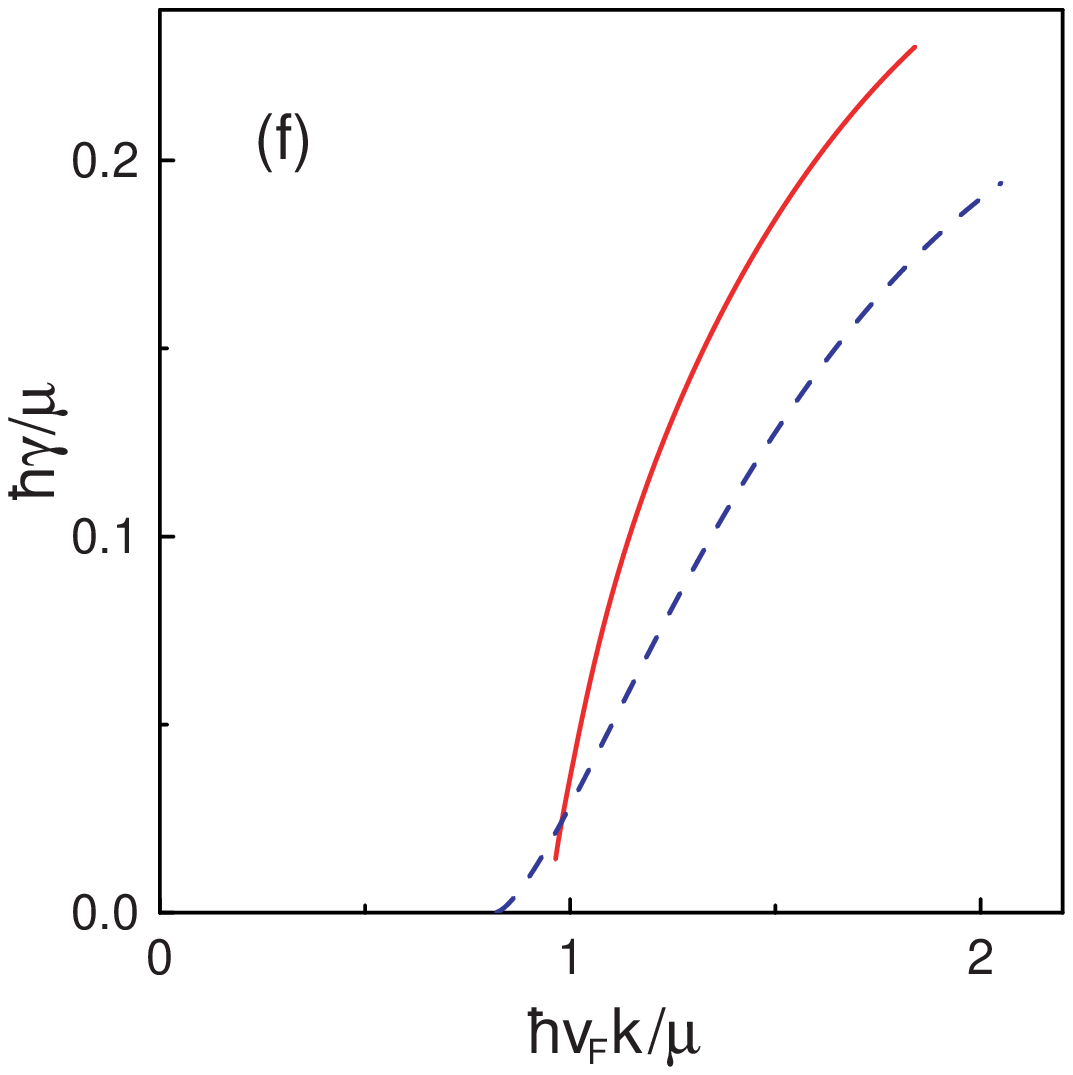}
\caption{Panels (a), (b), (c), and (d) show the plasmon spectrum for $\Delta=0$, $\Delta=0.2\mu$, $\Delta=0.3\mu$, and $\Delta=0.9\mu$, respectively, and (e) is a magnified fragment of (b). Solid lines represent the solutions of the complex dispersion equation, dotted lines are the solutions of the approximate equation~(\ref{realeqn}) and dashed curves show the long-wavelength square root dependence. The shaded area indicates the SPE regions in which plasmons decay with the damping rate shown in panel (f) for $\Delta=0$ (dashed line) and $\Delta=0.2\mu$ (solid line).}
\label{figure1}
\end{figure*}

Regions {\rm 1A}--{\rm 2A} and {\rm 2B}--{\rm 4B} (Fig.~\ref{figureA1}) in the ($k,\omega$) space, in which the imaginary part of the polarization function is different from zero, form the single-particle excitation (SPE) continuum, displayed in Fig.~\ref{figure1} as a shaded area. In those regions plasmons decay into electron-hole pairs (Landau damping) and equation~(\ref{dispeqn}) has no real solutions. Then one has to solve the complex dispersion equation
\begin{equation}
\Pi(\omega_p-i\gamma,k)+k=0\,,
\label{complexeqn}
\end{equation}
finding both energy $\hbar\omega_p(k)$ and the decay rate $\gamma(k)$ of plasmons. We solve~(\ref{complexeqn}) numerically, using analytical continuation of~(\ref{repi})--(\ref{impi}) from the real $\omega$ axis into the lower complex half-plane.

When the damping of the collective mode is weak, the approximate equation
\begin{equation}
\Re e\,\Pi(\omega_p,k)+k=0
\label{realeqn}
\end{equation}
is often solved~\cite{Wunsch2006NJP,Wang2007PRB} instead of~(\ref{complexeqn}) and the decay rate, which is assumed to be small, is determined from another equation,
\begin{equation}
\gamma=\frac{\Im m\,\Pi(\omega_p,k)}{(\partial/\partial\omega)\Re e\,\Pi(\omega,k)\bigr|_{\omega=\omega_p}}\,. 
\label{dampingrate}
\end{equation}
The numerical solutions of equations~(\ref{complexeqn}) (for the frequency and the damping rate) and~(\ref{realeqn}) together with the approximate analytical solution~(\ref{dispsmallk}) are plotted in Fig.~\ref{figure1} for different values of the gap parameter.

\begin{figure}
\includegraphics[width=6cm]{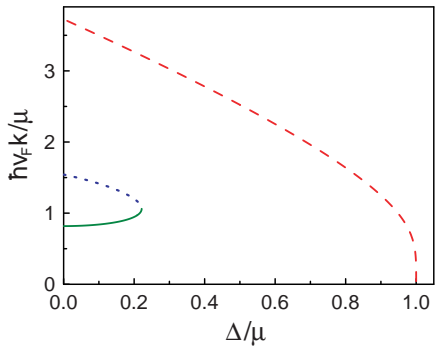}
\caption{Coordinates (momenta) of the undamped plasmon mode edges as a function of the gap value: the end of the mode starting at $k=0$ (solid line), the beginning (dotted line) and the end (dashed line) of the second mode which is absent in the gapless case.}
\label{figure2}
\end{figure}

At zero gap there is one plasmon mode which is undamped until it enters the SPE region (Fig.~\ref{figure1}(a))~\cite{Wunsch2006NJP,Hwang2007PRB}. When the gap is different from zero but is still small ($\Delta\lesssim0.22\mu$), a new undamped plasmon mode emerges in the gap which opens between two SPE regions (corresponding to intra- and interband transitions of electrons) in the vicinity of the line $\omega=v_Fk$, as displayed in Fig.~\ref{figure1}(b). The damping part of the plasmon spectrum is no longer a continuous extension of the undamped part but rather a separate mode (Fig.~\ref{figure1}(e)). When $\Delta$ becomes larger than $\simeq0.22\mu$ (this threshold value depends on the background dielectric constant, for example in the vacuum, when $\varepsilon_0=1$, it is approximately equal to $0.29\mu$), the damped part of the spectrum disappears and two undamped modes merge (Fig.~\ref{figure1}(c)). With the further increasing of $\Delta$, the plasmon dispersion curve becomes shorter (Fig.~\ref{figure1}(d)) and vanishes at $\Delta=\mu$. The gap dependence of the plasmon spectrum described above is also shown in Fig.~\ref{figure2}.

Thus we found that plasmons are absent at $\Delta\geq\mu$. Indeed, the dispersion equation~(\ref{dispeqn}) can be satisfied only when $\Pi_0(\omega,k)<0$, and it can be shown from~(\ref{repi0}) that $\Re e\,\Pi_0(\omega,k)>0$ in the region where $\Pi_0(\omega,k)$ is real (however, for the gapless case this conclusion was shown to be an artefact of the RPA approximation~\cite{Gangadharaiah2008PRB}, and the ladder corrections to polarization function lead to the existence of the undamped plasmon mode even at $\mu=\Delta=0$).

In those regions where the solutions of equations~(\ref{complexeqn}) and~(\ref{realeqn}) are close to each other, the decay rate given by~(\ref{dampingrate}) is small and does not differ significantly from $\gamma$ obtained from~(\ref{complexeqn}). When the solutions move away from each other, it turns out that the expression~(\ref{dampingrate}) is not small, and the approximate equations~(\ref{realeqn}) and~(\ref{dampingrate}) are not valid. In particular, the lower branch which appears for $\Delta\ne0$ and lies fully in the intraband SPE region (Figs.~\ref{figure1}(b)-(e)) is merely an artefact of this approximation.

Our numerical results are in agreement with the work~\cite{Wang2007PRB}, where the equation~(\ref{realeqn}) has been solved and a graph, similar to Fig.~\ref{figure1}(c) was obtained for the case of zero temperature.

\section{Conclusions}
We have calculated analytically the one-loop polarization function in the presence of a finite gap in the spectrum of Dirac quasiparticles. The result is expressed in terms of elementary functions and is valid for arbitrary wavevector, frequency, doping, and gap. The obtained expression is employed to calculate the static screening of Coulomb potential and the spectrum of plasmons within the RPA approximation. The oscillating part of the screened potential decays slower than in the gapless graphene, while the non-oscillating part is the same. It turns out that collective plasmon modes are present only at $\Delta<\mu$, i.e. when the Fermi level lies outside the gap. At long wavelengths the plasmon dispersion shows the usual square root dependence for 2D systems. At non-zero gap, an additional undamped mode is present at small values of $\Delta/\mu$, and at $0.22\mu\lesssim\Delta<\mu$ only one undamped plasmon mode remains.

\begin{acknowledgments}
The author is sincerely grateful to Prof. V.P.~Gusynin for the
formulation of the problem and many useful recommendations.
\end{acknowledgments}

\appendix*

\def\thefigure{A.\arabic{figure}}
\setcounter{figure}{0}

\section{Calculation of ${\bf \Pi}$($\boldsymbol\omega$,\,$\boldsymbol k$)}
To obtain the retarded polarization function we start with a finite temperature $T$. In the one-loop approximation it is given by the expression
\begin{figure}
\includegraphics[width=7.5cm]{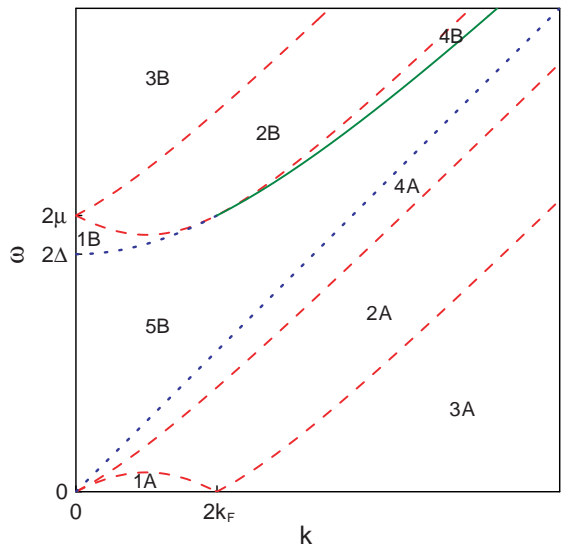}
\caption{Regions with different expressions for polarization function (here $\Delta=0.86\mu$). At the borders $\Pi(\omega,k)$ is smooth (dotted line) and has a logarithmic singularity (solid line) or discontinuity of its derivative (dashed line).}
\label{figureA1}
\end{figure}
\begin{equation}
\begin{split}
\Pi(&i\omega_m,k)=\frac{2\pi e^2TN_f}{\varepsilon_0}\sum_{n=-\infty}^{+\infty}\int\frac{d^2q}{(2\pi)^2}\\
&\times{\rm tr}\bigl[\gamma_0S(i\omega_m+i\Omega_n,{\bf k}+{\bf q})\gamma_0S(i\Omega_n,{\bf q})\bigr]\,,
\end{split}
\end{equation}
where $\omega_m=2\pi mT/\hbar$ and $\Omega_n=(2n+1)\pi T/\hbar$ are
Matsubara frequencies, and the propagator of Dirac quasiparticles has the form
\begin{equation}
S(i\Omega_n,{\bf q})=\frac{i}{(i\hbar\Omega_n-\mu)\gamma^0-\hbar
v_F{\bf q}{\boldsymbol\gamma}-\Delta}\,.
\end{equation}
Here we are not taking into account the spin-orbit gap, considering only the sublattice symmetry breaking mass term~(\ref{diracgap}). After taking the trace over spinor indices we obtain (for convenience we shall omit constants $\hbar$ and $v_F$ during intermediate calculations)
\begin{equation}
\begin{split}
&\Pi(i\omega_m,k)=-\frac{2e^2TN_f}{\pi\varepsilon_0}\sum_{n=-\infty}^{+\infty}\int d^2q\\
&\times\!\!\frac{(i\omega_m+i\Omega_n-\mu)(i\Omega_n-\mu)+{\bf q}({\bf k}+{\bf q})+\Delta^2}{\bigl[(i\omega_m\!+i\Omega_n-\mu)^2-E_{{\bf k}+{\bf q}}^2\bigr]\bigl[(i\Omega_n-\mu)^2-E_{\bf q}^2\bigr]},
\end{split}
\end{equation}
with $E_{{\bf q}}=\sqrt{q^2+\Delta^2}$. To perform the summation over $n$ we rewrite the integrand as
\begin{equation}
\begin{split}
\sum_{\lambda,\lambda'=\pm}&\frac14\biggl(1+\lambda\lambda'\frac{{\bf q}({\bf k}+{\bf q})+\Delta^2}{E_{\bf q}E_{{\bf k}+{\bf q}}}\biggr)\\
&\!\!\!\!\!\!\times\frac{1}{i\Omega_n-\mu+\lambda E_{\bf q}}\frac{1}{i\Omega_n+i\omega_m-\mu+\lambda'E_{{\bf k}+{\bf q}}}\,,
\end{split}
\end{equation}
where $\lambda$, $\lambda'$ are band indices, and then use the formula
\begin{equation}
\begin{split}
\sum_{n=-\infty}^{+\infty}\frac{1}{i\Omega_n-\mu+x_1}&\frac{1}{i\Omega_n+i\omega_m-\mu+x_2}\\
&=\frac1T \frac{n_F(x_1)-n_F(x_2)}{x_1-x_2-i\omega_m}\,,
\end{split}
\end{equation}
where $n_F(x)=\bigl[e^{(x-\mu)/T}+1\bigr]^{-1}$. After summation, the analytic continuation from Matsubara frequencies is made by the replacement $i\omega_m\to\omega+i0$, and the retarded polarization function reads
\begin{align}
\Pi(\omega,k)&=-\frac{e^2N_f}{2\pi\varepsilon_0}\int d^2q\sum_{\lambda,\lambda'=\pm}\biggl(1+\lambda\lambda'\frac{{\bf q}({\bf k}+{\bf q})+\Delta^2}{E_{\bf q}E_{{\bf k}+{\bf q}}}\biggr)\nonumber\\
&\times\frac{n_F(\lambda E_{\bf q})-n_F(\lambda'E_{{\bf k}+{\bf q}})}{\lambda E_{\bf q}-\lambda'E_{{\bf k}+{\bf q}}-\omega-i0}\,.
\label{piretarded}
\end{align}
The same expression (with $\Delta_{so}$ instead of $\Delta$) is obtained if the sublattice symmetry is unbroken and the spin-orbit gap~(\ref{spinorbitgap}) is taken into account~\cite{Wang2007PRB}, hence our results will be valid for both cases. At zero temperature Fermi functions $n_F(x)$ turn into step functions $\theta(\mu-x)$, and~(\ref{piretarded}) gives
\begin{equation}
\Pi(\omega,k)=-\chi_\infty^-(\omega,k)+\chi_\mu^+(\omega,k)+\chi_\mu^-(\omega,k)\,,
\label{pizerotemperature}
\end{equation}
where
\begin{align}
\chi_D^\pm&(\omega,k)=-\frac{e^2N_f}{2\pi\varepsilon_0}\int d^2q\, \theta(D^2-\Delta^2-q^2)\nonumber\\
&\times\biggl(1\pm\frac{{\bf q}({\bf k}+{\bf q})+\Delta^2}{E_{\bf q}E_{{\bf k}+{\bf q}}}\biggr)\label{chi}\\
&\times\biggl(\frac{1}{\omega+E_{\bf q}\mp E_{{\bf k}+{\bf q}}+i0}-\frac{1}{\omega-E_{\bf q}\pm E_{{\bf k}+{\bf q}}+i0}\biggr)\,.\nonumber
\end{align}
Here the upper and the lower signs correspond to intraband and interband electron-hole transitions, respectively, and the parameter $D$ defines the integration limits. We proceed with further calculations in a similar way to those shown in~\cite{Wunsch2006NJP} for massless Dirac quasiparticles ($\Delta=0$). In the case $\mu<\Delta$ (when the Fermi level lies in the band gap) only the first term in~(\ref{pizerotemperature}) survives. Thus we can split our polarization function into two parts
\begin{equation}
\Pi(\omega,k)=\Pi_0(\omega,k)+\theta(\mu-\Delta)\,\Pi_1(\omega,k)\,,
\end{equation}
where $\Pi_0(\omega,k)=-\chi_\infty^-(\omega,k)$ is the polarization function at chemical potential $\mu<\Delta$, which does not depend on the value of $\mu$, and
\begin{equation}
\Pi_1(\omega,k)=\chi_\mu^+(\omega,k)+\chi_\mu^-(\omega,k)
\label{pichipm}
\end{equation}
is a correction to the polarization function in the case when the chemical potential is greater than the gap parameter $\Delta$.

At first we consider the imaginary part of~(\ref{chi}),
\begin{align}
\Im
m\,&\chi_D^\pm(\omega,k)=\frac{e^2N_f}{2\varepsilon_0}\,\theta(D-\Delta)\int\limits_0^{\lefteqn{\scriptstyle\sqrt{D^2-\Delta^2}}}d q\,q\int\limits_0^{2\pi}d\varphi\nonumber\\
&\times\biggl(1\pm\frac{{\bf q}({\bf k}+{\bf q})+\Delta^2}{E_{\bf q}E_{{\bf k}+{\bf q}}}\biggr)\\
&\times\Bigl[\delta(\omega+E_{\bf q}\mp E_{{\bf k}+{\bf q}})-\delta(\omega-E_{\bf q}\pm E_{{\bf k}+{\bf q}})\Bigr]\,.\nonumber
\end{align}
Arguments of these $\delta$-functions determine the SPE regions in the ($k,\omega$) space. The integration over $\varphi$ yields
\begin{align}
\Im m\,&\chi_D^\pm(\omega,k)=\frac{e^2N_f}{\varepsilon_0}\,\theta(D-\Delta)\sum_{\beta=\pm}\int\limits_\Delta^Dd E\nonumber\\
&\times\Bigl[\theta(\beta)\theta(\pm1)\mp\theta(-\beta)\theta(\pm E\mp\omega)\Bigr]\nonumber\\
&\times\Bigl[(2E+\beta\omega)^2-k^2\Bigr]
\label{imchi}\\
&\times\frac{\theta\bigl((k^2-\omega^2)\bigl[(2E+\beta\omega)^2-k^2\bigr]-4k^2\Delta^2\bigr)}{\sqrt{(k^2-\omega^2)\bigl[(2E+\beta\omega)^2-k^2\bigr]-4k^2\Delta^2}}\,.\nonumber
\end{align}
Taking the lower sign and setting $D=\infty$ in the above expression and integrating over $E$ we obtain the imaginary part of polarization at $\mu<\Delta$,
\begin{align}
\Im m\,\Pi_0(&\omega,k)=-\Im m\,\chi_\infty^-(\omega,k)=\frac{\pi e^2N_fk^2}{4\varepsilon_0\sqrt{\omega^2-k^2}}\nonumber\\
&\times\theta(\omega^2-k^2-4\Delta^2)\biggl(1+\frac{4\Delta^2}{\omega^2-k^2}\biggr)\,.
\label{impi0}
\end{align}
We find the real part of the polarization function at $\mu<\Delta$ from the Kramers-Kronig relation, using its already known imaginary part~(\ref{impi0}), instead of direct calculation of $\Re e\,\chi_\infty^-(\omega,k)$ from equation~(\ref{chi})
\begin{equation}
\begin{split}
\Re e\,&\Pi_0(\omega,k)=\frac1\pi{\,\rm v.p.\!\!}\int\limits_{-\infty}^\infty d\omega'\frac{\Im m\,\Pi_0(\omega',k)}{\omega'-\omega}\\
&=\frac{e^2N_fk^2}{4\varepsilon_0}{\,\rm v.p.\!\!}\int\limits_{-\infty}^\infty d\omega'\frac{\theta(\omega'^2-k^2-4\Delta^2)}{(\omega'-\omega)\sqrt{\omega'^2-k^2}}\\
&\times\biggl(1+\frac{4\Delta^2}{\omega'^2-k^2}\biggr){\rm sgn\,}\omega'\,.
\end{split}
\end{equation}
After integration we have
\begin{align}
\Re e\,&\Pi_0(\omega,k)=\frac{e^2N_fk^2}{\varepsilon_0}\biggl\{\frac{\Delta}{k^2-\omega^2}\nonumber\\
&+\frac{k^2-\omega^2-4\Delta^2}{4|k^2-\omega^2|^{3/2}}\biggl[\theta(k-\omega)\arccos\frac{k^2-\omega^2-4\Delta^2}{\omega^2-k^2-4\Delta^2}\nonumber\\
&-\theta(\omega-k)\ln\frac{(2\Delta+\sqrt{\omega^2-k^2})^2}{|\omega^2-k^2-4\Delta^2|}\biggr]\biggr\}\,.
\label{repi0}
\end{align}
We can combine~(\ref{impi0}) and~(\ref{repi0}), obtaining the
retarded version of the time component of the well-known expression
for the polarization tensor in QED$_3$~\cite{Appelquist1986PRD}
\begin{align}
\Pi_0(\omega,k)&=\frac{e^2N_fk^2}{2\varepsilon_0\hbar(v_F^2k^2\!-\omega^2)}\biggl(\frac{2\Delta}{\hbar}+\frac{v_F^2k^2\!-\omega^2\!-4\Delta^2\!/\hbar^2}{\sqrt{v_F^2k^2-\omega^2}}\nonumber\\
&\times\arcsin\sqrt{\frac{v_F^2k^2-\omega^2}{v_F^2k^2-\omega^2+4\Delta^2/\hbar^2}}\;\biggr)\,,
\label{pizeromu}
\end{align}
where the prescription $\omega\to\omega+i0$ is used (compare also
with equation (6) in~\cite{Kotov2008PRB}).

Using the expressions~(\ref{pichipm}) and~(\ref{imchi}) we can write the correction to the imaginary part of the polarization function as
\begin{align}
\Im m\,&\Pi_1(\omega,k)=\frac{e^2N_f}{\varepsilon_0}\sum_{\beta=\pm} \int_\Delta^\mu dE\nonumber\\
&\times{\,\rm sgn}(\omega+\beta E)\Bigl[(2E+\beta\omega)^2-k^2\Bigr]
\label{impi1}\\
&\times\frac{\theta\bigl((k^2-\omega^2)\bigl[(2E+\beta\omega)^2-k^2\bigr]-4k^2\Delta^2\bigr)}{\sqrt{(k^2-\omega^2)\bigl[(2E+\beta\omega)^2-k^2\bigr]-4k^2\Delta^2}}\,.\nonumber
\end{align}
The real part of $\Pi_1(\omega,k)$ is obtained from~(\ref{pichipm}) and~(\ref{chi}), and after some algebraic simplifications it reads
\begin{equation}
\begin{split}
\Re e\,&\Pi_1(\omega,k)=\frac{e^2N_f}{\pi\varepsilon_0}\int_0^{\sqrt{\mu^2-\Delta^2}}\frac{dq\,q}{E_{\bf q}}{\,\rm v.p.\!\!}\int_0^{2\pi}d\varphi\\
&\times\biggl[\frac{(\omega+2E_{\bf q})E_{\bf q}+{\bf qk}}{k^2-\omega(\omega+2E_{\bf q})+2{\bf qk}}+(\omega\to-\omega)\biggr]\,.
\end{split}
\end{equation}
Integrating this expression over the angle we find
\begin{align}
\Re e&\,\Pi_1(\omega,k)=\frac{e^2N_f}{\varepsilon_0}\biggl\{\int_\Delta^\mu dE\biggl[{\,\rm sgn}\bigl(k^2-\omega(2E+\omega)\bigr)\nonumber\\
&\times\frac{\theta\bigl((\omega^2-k^2)\bigl[(2E+\omega)^2-k^2\bigr]+4k^2\Delta^2\bigr)}{\sqrt{(\omega^2-k^2)\bigl[(2E+\omega)^2-k^2\bigr]+4k^2\Delta^2}}
\label{repi1}\\
&\times\bigl[(2E+\omega)^2-k^2\bigr]+(\omega\to-\omega)\biggr]+2(\mu-\Delta)\biggr\}\,.\nonumber
\end{align}
Performing the last integration in~(\ref{impi1}) and~(\ref{repi1}) and adding the results to~(\ref{impi0}) and~(\ref{repi0}), respectively, we arrive at the final expression for the polarization function at $\mu>\Delta$. Restoring constants $\hbar$ and $v_F$ and introducing the following notations
\begin{align}
f(\omega,k)&=\frac{e^2N_fk^2}{4\varepsilon_0\hbar\sqrt{|v_F^2k^2-\omega^2|}}\,,\nonumber\\
x_0&=\sqrt{1+\frac{4\Delta^2}{\hbar^2(v^2_Fk^2-\omega^2)}}\,,\nonumber\\
G_<(x)&=x\sqrt{x_0^2-x^2}-(2-x_0^2)\arccos(x/x_0)\,,\nonumber\\
G_>(x)&=x\sqrt{x^2-x_0^2}-(2-x_0^2){\,\rm arccosh}(x/x_0)\,,\nonumber\\
G_0(x)&=x\sqrt{x^2-x_0^2}-(2-x_0^2){\,\rm arcsinh}\Bigl(x/\sqrt{-x_0^2}\Bigr)\,,\nonumber
\end{align}
we can write the results in the next form
\begin{equation}
\begin{split}
\Re e&\,\Pi(\omega, k)=\frac{2e^2N_f\mu}{\varepsilon_0\hbar^2v^2_F}-f(\omega,k)\\
&\times\left\{
\begin{array}{ll}
0\,,&{\rm1A}\\
\displaystyle G_<\Bigl(\frac{2\mu-\hbar\omega}{\hbar v_Fk}\Bigr)\,,&{\rm2A}\medskip\\
\displaystyle G_<\Bigl(\frac{2\mu+\hbar\omega}{\hbar v_Fk}\Bigr)+G_<\Bigl(\frac{2\mu-\hbar\omega}{\hbar v_Fk}\Bigr)\,,\;&{\rm3A}\medskip\\
\displaystyle G_<\Bigl(\frac{2\mu-\hbar\omega}{\hbar v_Fk}\Bigr)-G_<\Bigl(\frac{2\mu+\hbar\omega}{\hbar v_Fk}\Bigr)\,,&{\rm4A}\medskip\\
\displaystyle G_>\Bigl(\frac{2\mu+\hbar\omega}{\hbar v_Fk}\Bigr)-G_>\Bigl(\frac{2\mu-\hbar\omega}{\hbar v_Fk}\Bigr)\,,&{\rm1B}\medskip\\
\displaystyle G_>\Bigl(\frac{2\mu+\hbar\omega}{\hbar v_Fk}\Bigr)\,,&{\rm2B}\medskip\\
\displaystyle G_>\Bigl(\frac{2\mu+\hbar\omega}{\hbar v_Fk}\Bigr)-G_>\Bigl(\frac{\hbar\omega-2\mu}{\hbar v_Fk}\Bigr)\,,&{\rm3B}\medskip\\
\displaystyle G_>\Bigl(\frac{\hbar\omega-2\mu}{\hbar v_Fk}\Bigr)+G_>\Bigl(\frac{2\mu+\hbar\omega}{\hbar v_Fk}\Bigr)\,,&{\rm4B}\medskip\\
\displaystyle G_0\Bigl(\frac{2\mu+\hbar\omega}{\hbar v_Fk}\Bigr)-G_0\Bigl(\frac{2\mu-\hbar\omega}{\hbar v_Fk}\Bigr)\,,&{\rm5B}
\end{array}
\right.
\label{repi}
\end{split}
\end{equation}
\begin{equation}
\begin{split}
\Im m&\,\Pi(\omega, k)=f(\omega,k)\\
&\!\!\!\times\left\{
\begin{array}{ll}
\displaystyle G_>\Bigl(\frac{2\mu+\hbar\omega}{\hbar v_Fk}\Bigr)-G_>\Bigl(\frac{2\mu-\hbar\omega}{\hbar v_Fk}\Bigr)\,,\;&{\rm1A}\medskip\\
\displaystyle G_>\Bigl(\frac{2\mu+\hbar\omega}{\hbar v_Fk}\Bigr)\,,&{\rm2A}\\
0\,,&{\rm3A}\\
0\,,&{\rm4A}\\
0\,,&{\rm1B}\\
\displaystyle{}-G_<\Bigl(\frac{2\mu-\hbar\omega}{\hbar v_Fk}\Bigr)\,,&{\rm2B}\medskip\\
\pi(2-x_0^2)\,,&{\rm3B}\smallskip\\
\pi(2-x_0^2)\,,&{\rm4B}\\
0\,,&{\rm5B}\\
\end{array}
\right.
\label{impi}
\end{split}
\end{equation}
with the following regions in the ($k,\omega$) space (shown in Fig.~\ref{figureA1})
\begin{equation}
\begin{array}{ll}
{\rm1A :}\;&\hbar\omega<\mu-\sqrt{\hbar^2v_F^2(k-k_F)^2+\Delta^2}\,,\medskip\\
{\rm2A :}&\pm\mu\mp\sqrt{\hbar^2v_F^2(k-k_F)^2+\Delta^2}<\hbar\omega\smallskip\\
&<-\mu+\sqrt{\hbar^2v_F^2(k+k_F)^2+\Delta^2}\,,\medskip\\
{\rm3A :}&\hbar\omega<-\mu+\sqrt{\hbar^2v_F^2(k-k_F)^2+\Delta^2}\,,\medskip\\
{\rm4A :}&-\mu+\sqrt{\hbar^2v_F^2(k+k_F)^2+\Delta^2}<\hbar\omega<\hbar v_Fk\,,\medskip\\
{\rm1B :}&k<2k_F\,,\qquad\sqrt{\hbar^2v^2_Fk^2+4\Delta^2}< \hbar\omega\smallskip\\
&<\mu+\sqrt{\hbar^2v_F^2(k-k_F)^2+\Delta^2}\,,
\end{array}
\nonumber
\end{equation}
\begin{equation}
\begin{array}{ll}
{\rm2B :}\;&\mu+\sqrt{\hbar^2v_F^2(k-k_F)^2+\Delta^2}<\hbar\omega\smallskip\\
&<\mu+\sqrt{\hbar^2v_F^2(k+k_F)^2+\Delta^2}\,,\medskip\\
{\rm3B :}&\hbar\omega>\mu+\sqrt{\hbar^2v_F^2(k+k_F)^2+\Delta^2}\,,\medskip\\
{\rm4B :}&k>2k_F\,,\qquad\sqrt{\hbar^2v^2_Fk^2+4\Delta^2}<\hbar\omega\smallskip\\
&<\mu+\sqrt{\hbar^2v_F^2(k-k_F)^2+\Delta^2}\,,
\end{array}
\nonumber
\end{equation}
\newline
$
~~~~~{\rm5B :}\quad\hbar v_Fk<\hbar\omega<\sqrt{\hbar^2v^2_Fk^2+4\Delta^2}\,,
$

\noindent where $\displaystyle k_F\equiv\sqrt{\mu^2-\Delta^2}/\hbar v_F$.

In the gapless limit $x_0$ becomes equal to 1, regions 4A, 4B, and 5B vanish, and the expressions~(\ref{repi})--(\ref{impi}) reduce to~(\ref{pizerodelta}).

\


\begin{thebibliography}{99}

\bibitem{Novoselov2004S} K.S.~Novoselov, A.K.~Geim, S.V.~Morozov, D.~Jiang, Y.~Zhang, S.V.~Dubonos, I.V.~Grigorieva and A.A.~Firsov, Science {\bf306}, 666 (2004);
K.S.~Novoselov, A.K.~Geim, S.V.~Morozov, D.~Jiang, M.I.~Katsnelson, I.V.~Grigorieva, S.V.~Dubonos and A.A.~Firsov, Nature {\bf438}, 197 (2005);
Y.~Zhang, Y.-W.~Tan, H.L.~Stormer and P.~Kim, Nature {\bf438}, 201 (2005).

\bibitem{Wallace1947PR} P.R.~Wallace, Phys. Rev. {\bf71}, 622 (1947).

\bibitem{DiVincenzo1984PRB} D.P.~DiVincenzo and E.J.~Mele, Phys. Rev. B {\bf29}, 1685 (1984); G.W.~Semenoff, Phys. Rev. Lett. {\bf53}, 2449 (1984).

\bibitem{Gorbar2002PRB} E.V.~Gorbar, V.P.~Gusynin, V.A.~Miransky and I.A.~Shovkovy, Phys. Rev. B {\bf66}, 045108 (2002).

\bibitem{Wunsch2006NJP} B.~Wunsch, T.~Stauber, F.~Sols and F.~Guinea, New~J.~Phys. {\bf8}, 318 (2006).

\bibitem{Hwang2007PRB} E.H.~Hwang and S.~Das~Sarma, Phys. Rev. B {\bf75}, 205418 (2007).

\bibitem{Zhou2007NM} S.Y.~Zhou, G.-H.~Gweon, A.V.~Fedorov, P.N.~First, W.A.~de~Heer, D.-H.~Lee, F.~Guinea, A.H.~Castro Neto and A.~Lanzara, Nat. Mater. {\bf6}, 770 (2007).

\bibitem{Li2008} G.~Li, A.~Luican and E.Y.~Andrei, arXiv:0803.4016.

\bibitem{Huertas-Hernando2006PRB} D.~Huertas-Hernando, F.~Guinea and A.~Brataas, Phys. Rev. B {\bf74}, 155426 (2006);
H.~Min, J.E.~Hill, N.A.~Sinitsyn, B.R.~Sahu, L.~Kleinman and A.H.~MacDonald, Phys.~Rev. B {\bf74}, 165310 (2006);
Y.~Yao, F.~Ye, X.-L.~Qi, S.-C.~Zhang and Z.~Fang, Phys. Rev. B {\bf75}, 041401 (2007).

\bibitem{Khveshchenko2001PRL} D.V.~Khveshchenko, Phys. Rev. Lett. {\bf87}, 206401 (2001);
E.V.~Gorbar, V.P.~Gusynin, V.A.~Miransky and I.A.~Shovkovy, Phys. Rev. B {\bf78}, 085437 (2008).

\bibitem{Wang2007PRB} X.F.~Wang and T.~Chakraborty, Phys. Rev. B {\bf75}, 033408 (2007).

\bibitem{Kane2005PRL} C.L.~Kane and E.J.~Mele, Phys. Rev. Lett. {\bf95}, 226801 (2005).

\bibitem{Barlas2007PRL} Y.~Barlas, T.~Pereg-Barnea, M.~Polini, R.~Asgari and A.H.~MacDonald, Phys. Rev. Lett. {\bf98}, 236601 (2007).

\bibitem{Stern1967PRL} F.~Stern, Phys. Rev. Lett. {\bf18}, 546 (1967).

\bibitem{Cheianov2006PRL} V.V.~Cheianov and V.I.~Fal'ko, Phys. Rev. Lett. {\bf97}, 226801 (2006).

\bibitem{Kotov2008PRB} V.N.~Kotov, V.M.~Pereira and B.~Uchoa, Phys. Rev. B {\bf78}, 075433 (2008).

\bibitem{Gangadharaiah2008PRB} S.~Gangadharaiah, A.M.~Farid and E.G.~Mishchenko, Phys. Rev. Lett. {\bf100}, 166802 (2008).

\bibitem{Appelquist1986PRD} T.W.~Appelquist, M.~Bowick, D.~Karabali and L.C.R.~Wijewardhana, Phys. Rev. D {\bf33}, 3704 (1986).

\end{thebibliography}
\end{document}